\documentclass[12pt]{iopart}
\usepackage{graphicx}
\usepackage{amssymb}
\usepackage{color}
\newcommand{\rhoml}{\rho_{\mathrm{ML}}}
\newcommand{\rhotrue}{\rho_{\mathrm{true}}}
\newcommand{\setx}{\{x_i\}}

\begin{document}
\title{ Gradient-based stopping rules for maximum-likelihood
  quantum-state tomography}

\author{ S Glancy, E Knill and M Girard\footnote{Present address:
    Freiburg Institute for Advanced Studies,
    Albert-Ludwigs-Universit\"{a}t, Albertstrasse 19, D-79104
    Freiburg, Germany} }

\address{Applied and Computational Mathematics Division, National
  Institute of Standards and Technology, Boulder, Colorado, USA}
\ead{sglancy@nist.gov}

\begin{abstract}
When performing maximum-likelihood quantum-state tomography, one must
find the quantum state that maximizes the likelihood of the state
given observed measurements on identically prepared systems.  The
optimization is usually performed with iterative algorithms.  This
paper provides a gradient-based upper bound on the ratio of the true
maximum likelihood and the likelihood of the state of the current
iteration, regardless of the particular algorithm used.  This bound is
useful for formulating stopping rules for halting iterations of
maximization algorithms.  We discuss such stopping rules in the
context of determining confidence regions from log-likelihood
differences when the differences are approximately chi-squared
distributed.

\end{abstract}

\section{Introduction}

Quantum-state tomography is a statistical procedure for estimating the
quantum state $\rho_\mathrm{true}$ of a quantum system.  One prepares
many identical copies of the system, measures each copy independently
and uses the measurement results to estimate $\rho_\mathrm{true}$.
One useful way to make the estimate is to find the state
$\rho_{\mathrm{ML}}$ with the maximum likelihood for the measurement
results \cite{hradil2004}.  The estimation then becomes an
optimization problem, which is usually solved numerically with
iterative methods.  A stopping rule is needed to decide when to halt
the iterations, otherwise one risks stopping before the calculation
has reached a point `near enough' to $\rho_{\mathrm{ML}}$ or wasting
time with unnecessary iterations.  Using the difference between the
state or the likelihood achieved in successive iterations is
unreliable, especially if the maximization algorithm suffers from slow
convergence.  Ideally, the stopping rule should specify `near enough'
in a statistically relevant way: if there is large statistical
uncertainty, high numerical precision may not be necessary.  In this
paper we give such stopping rules that depend on an upper bound on the
ratio of the true maximum likelihood and the likelihood of the state
achieved at the current iteration.  The bound and stopping rules can
be applied to any iterative likelihood maximization algorithm.  The
bound is particularly useful when a likelihood ratio is used to assign
uncertainties to the inferred state.

This paper begins with the derivation of a bound on the ratio of the
true maximum likelihood to the likelihood of a particular state.  One
may stop iterations when this ratio is sufficiently small.  We next
give a brief review of Wilks's Theorem, which gives the probability
distribution for the likelihood ratio statistic.  We use this theorem
to give some rules-of-thumb for when one should halt iterations, in
three contexts: (1) Point estimation.  The goal is to obtain a point
estimate for which the likelihood of the estimate is at least as large
as the expectation value of the likelihood of the true state with respect to
the data.  (2) Confidence regions for states.  Here the goal is to
construct a confidence region for the true state at a given
significance level. (3) Confidence regions for expectation values.  In
this case, we wish to construct a confidence region for expectation
values of observables of the state.  Our stopping criteria are
formulated using Wilks's Theorem, which may not always apply in
quantum state tomography experiments.  We then present a numerical
example of our likelihood ratio bound using simulated optical homodyne
measurements.

\section{Likelihood ratio bound}
Suppose $N$ quantum systems are prepared, each in the state with
density matrix $\rho_{\mathrm{true}}$.  For each copy $i$, the
experimenter chooses an observable to measure.  We will label each
observable with $\theta_i$ for $i=1 \dots N$.  The measurements yield
results $x_i$.  Corresponding to the choice and result combination,
there is a positive-operator-valued-measure element
$\Pi\left(x_i|\theta_i\right) = \Pi_i$.  For finite-dimensional
systems, $N$ may exceed the number of possible measurement choice and
result combinations, so many of the $\Pi_i$ will equal one another.
However, for infinite-dimensional systems, the possible measurement
results may be infinite and continuous, so the $\Pi_i$ may all be
different.  The probability to observe $x$ when measuring $\theta$ is
$p(x|\theta) =
\Tr\left[\rho_{\mathrm{true}}\Pi\left(x|\theta\right)\right]$.  The
likelihood for observing the sequence of measurement results
$\{x_i:i=1\dots N\}$ as a function of candidate density matrix $\rho$
is
\[
\mathcal{L}\left(\rho\right)=\prod_{i=1}^{N}\Tr\left(\Pi_i\rho\right).
\]
The goal of maximum-likelihood quantum-state tomography is to maximize
this function to obtain the state $\rhoml$ as the estimate
of the true state $\rho_{\mathrm{true}}$.  The optimization is
easier if we focus instead on the natural logarithm of the likelihood
(the `log-likelihood'):
\[
L\left(\rho\right) = \ln\mathcal{L}\left(\rho\right) = \sum_{i=1}^{N}\ln\left[\Tr\left(\Pi_i\rho\right)\right].
\]
The same density matrix maximizes both the likelihood and the
log-likelihood.  Fortunately, the log-likelihood is concave over the
(convex) set of density matrices.  One can show that it is concave by
using the concavity of the logarithm, the linear dependence of the
individual event probabilities on the density matrix, and the fact
that the total log-likelihood is a positive sum of logarithms of the
probabilities.  This concavity simplifies
the maximization.  Several maximization methods are described in
Refs.~\cite{hradil2004, james2001, rehacek2007, goncalves2011}.  These
methods use iterative schemes, producing a new density matrix $\rho_k$
after the $k$'th iteration.

After iteration $k$, we would like to place an upper bound on
$L(\rhoml)-L(\rho_k)$. Consider the density matrix $\rho_\epsilon
=(1-\epsilon)\rho_k + \epsilon \rhoml$, where $0 \leq \epsilon \leq
1$.  Because the log-likelihood is concave, for any choice of
$\epsilon$
\[
L(\rho_\epsilon)-L(\rho_k) \leq \epsilon \left[\frac{\mathrm{d}
  L(\rho_\epsilon)}{\mathrm{d} \epsilon}\right]_{\epsilon=0}.
\]
In particular, when $\epsilon=1$,
\[
L(\rhoml)-L(\rho_k) \leq \left[\frac{\mathrm{d}
  L(\rho_\epsilon)}{\mathrm{d} \epsilon}\right]_{\epsilon=0}.
\]
The derivative evaluated at $\epsilon = 0$ is
\[
\left.\frac{\mathrm{d} L(\rho_\epsilon)}{\mathrm{d}
  \epsilon}\right|_{\epsilon=0} = \Tr[\rhoml R(\rho_k)]-N,
\]
where
\[
R(\rho_k)=\sum_{i=1}^{N} \frac{\Pi_i}{\Tr(\rho_k\Pi_i)}.
\]
This is the same matrix $R$ that is used in the `$R\rho R$' algorithm
described in \cite{rehacek2007}.  Of course, we do not know $\rhoml$,
so we find an upper bound of $\Tr[\rhoml R(\rho_k)]$ by maximizing $\Tr[\sigma
  R(\rho_k)]$ over all density matrices $\sigma$:
\[
L(\rhoml)-L(\rho_k) \leq \max_\sigma\Tr[\sigma R(\rho_k)]-N.
\]
This maximum is achieved for $\sigma$ equal to the pure density matrix
corresponding to the eigenstate of $R(\rho_k)$ with the largest eigenvalue.
Thus
\[
L(\rhoml)-L(\rho_k) \leq r(\rho_k),
\]
where $r(\rho_k)=r_k=\max\{\mathrm{eig}[R(\rho_k)]\}-N$.
After exponentiation, we obtain,
\[
\frac{\mathcal{L}(\rhoml)}{\mathcal{L}(\rho_{k})} \leq
e^{r_k}.
\]
Thus one may stop iterations when $r_k$ is less than a predetermined
bound specified by a stopping rule.  Specific bounds depend on context
as we discuss below.

The above ideas could also be adapted for a simple gradient-ascent
maximization procedure, as follows: Initialize the procedure with some
state $\rho_0$, perhaps the fully mixed state.  At each iteration, set
$\sigma$ equal to the eigenstate of $R(\rho_k)$ with the largest
eigenvalue.  Then use a one-dimensional optimization procedure to find
the $\epsilon$ maximizing $L(\rho_\epsilon)$ and set
$\rho_{k+1}=\rho_\epsilon$.  However, such a procedure can have slow
convergence because it uses only the slope of the log-likelihood
function and not its curvature.

\section{Review of Wilks's theorem}

Of course, the stopping rule, that is the value of $r_k$ below which
one can halt iterations, depends on how the estimate is used.  In the
following we discuss the use of $L(\rho_k)$ and $r_k$ to establish two
types of confidence regions related to our estimate.  The asymptotic
theory of likelihood-ratio tests provides guidelines.  A key technique
is the application of Wilks's Theorem; see Ref.~\cite{wilks1938} and
section 6.4 of Ref.~\cite{shao1998}.  Wilks's Theorem states that
under appropriate assumptions, for two sets of models $H_0\subseteq H$
specified by $h_0$ and $h$ free parameters, respectively, the random
variable $D(H_0|X) = 2[L(H_\mathrm{ML}|X)-L(H_{0,\mathrm{ML}}|X)]$
converges in distribution to $\chi^2(h-h_0)$, the chi-squared
distribution with $h-h_0$ degrees of freedom.  Here,
$L(H_{0,\mathrm{ML}}|X)$ and $L(H_\mathrm{ML}|X)$ are the maximum
log-likelihoods for $H_0$ and $H$, respectively, and we assume that
the true state is in the interior of $H_0$ with respect to the
parametrization. The parametrization must be sufficiently
well-behaved; see the references above. We can apply this to
hypotheses consisting of linear spaces of density matrices
parametrized with respect to a linear basis, provided the true density
matrix is not too near the boundary, that is, has no statistically
near-zero eigenvalues.

\section{Point estimate stopping rule}

As a first approximation to be refined below, we intuit that little
further information about $\rhotrue$ is obtained once
$L(\rhoml|\{x_i\}) - L(\rho_k|\{x_i\})$ is below $\left\langle
L(\rhoml|X) - L(\rhotrue|X)\right\rangle$, where $\langle . \rangle$
is the expectation value for the enclosed random variable, and $X$ is
a random vector of length $N$ distributed according to $\rhotrue$. If
$\rhotrue$ is an interior point of the space of density matrices and
$N$ is sufficiently large, the expectation of
$L(\rhoml|X)-L(\rhotrue|X)$ can be approximated by an application of
Wilks's Theorem.  That is, let $H$ consist of all density
matrices of dimension $d$; $H$ has $d^2-1$ free parameters.  Let $H_0$
contain only one element, $\rhotrue$. Then the random variable
$D(\rhotrue|X) = 2[L(\rhoml|X)-L(\rhotrue|X)]$ converges in
distribution to $\chi^2(d^2-1)$.  This distribution has expectation
$d^2-1$, so
\[
\left\langle L(\rhoml|X) - L(\rhotrue|X) \right\rangle = \frac{1}{2}(d^2-1).
\]
According to this intuition, one can stop iterations when $r_k$ is
less than a fraction of $(d^2-1)/2$.

To make the above intuition more precise, a reasonable stopping rule
can be based on the requirement that $\rho_k$ be in a confidence
region for the true state at a reasonable level of significance $s$.
Such a confidence region can be constructed from likelihood-ratio
hypothesis tests with level of significance $s$.  This confidence
region is defined as the set of density matrices (or other parameters)
$\rho$ for which the observations $\{x_i\}$ and the associated
likelihood ratio would not lead us to reject the hypothesis that
$\rho$ is $\rhotrue$ at level of significance $s$; see theorem 7.2 in
ref.~\cite{shao1998}. Here we reject the hypothesis that $\rho$ is
$\rhotrue$ if the observed log-likelihood difference $D(\rho|\{x_i\})$ has a
p-value less than $s$, where the p-value is the probability that the
state would (if it were the true state) produce a value for
$D(\rho|X)$ at least the observed value. In general, p-values are
associated with a random variable, in this case $D(\rho|X)$. For
brevity, we omit mention of the random variable when the random
variable is clear from context.  According to Wilks's Theorem, we can
calculate a state's p-value as the integral
\[
\textrm{p-value}=\int_{D(\rho|\setx)}^\infty f(u)\mathrm{d}u,
\]
where $f(u)$ is the probability density function for $\chi^2(d^2-1)$.
Notice that smaller values for $D(\rho|\setx)$ correspond to larger
p-values.  Let $t$ be the value of $D(\rho|\setx)$ that gives a
p-value equal to $s$. That is, $s=\int_t^\infty f(u)\mathrm{d}u$.
Thus the confidence region at level of significance $s$ is
$\{\rho:D(\rho|\{x_{i}\}) \leq t\}$.  We can ensure that our estimate
$\rho_k$ is in a confidence region at a predetermined level of
significance by stopping when $r_k$ is sufficiently small.  The level
of significance determines the statistical closeness of $\rho_k$ to
$\rhotrue$. Higher levels of significance imply closer $\rho_k$. For
example, a level of significance of $0.5$ requires that $r_k$ is at
most the median of the $\chi^2(d^2-1)$ distribution.  The mean and
variance of $\chi^2(f)$ distribution are $f$ and $2f$, respectively,
and as $f$ increases, $\chi^2(f)$ converges in distribution to a
Gaussian with the given mean and variance~\cite{dykstra1991}.
Therefore, to ensure that $\rho_k$ is in a confidence region for the
true state with $s\sim 0.5$, for large $d$, one may stop iterations
when $r_k$ is below $(d^2-1)/2$.

\section{State confidence region stopping rule}

Another potential use of $\rho_k$ is to construct a confidence region
of states based on $L(\rho_k|\setx)$. When determining confidence
regions rather than a statistically good approximation of $\rhotrue$,
it is not enough to ensure that $\rho_k$ is statistically close to
$\rhoml$.  Because $\rhoml$ is not known exactly, we construct a
confidence region at level of significance $s$ by replacing
$L(\rhoml|\setx)$ in the conventional definition of the
likelihood-ratio confidence region with the log-likelihood of our
estimate $L(\rho_k|\setx)$.  As we explain below, this confidence
region contains the conventional one.  For the confidence region to be
a good approximation of the maximum-likelihood confidence region
requires that $r_k$ is less than a fraction of $\sqrt{(d^2-1)/2}$, the
standard deviation of $\chi^2(d^2-1)$. This rule ensures that
approximate p-values computed according to
$2[L(\rho_k|\{x_i\})-L(\rho|\{x_i\})]$ and the actual p-values
computed with $D(\rho|\{x_i\})$ are sufficiently close.  As a
numerical example, consider tomography of a $d=10$ quantum system,
where we wish to construct the confidence region at level of
significance $0.32$.  In this case, $\sqrt{(d^2-1)/2}=7.04$, and the
threshold for $D(\rho|\{x_i\})$ is $t=105.04$. Suppose we stop
iterations when $r_k\precsim 2$.  If we construct a confidence region
as the set of $\rho$ for which $2[L(\rho_k|\{x_i\})-L(\rho|\{x_i\})]
\leq t$, the region includes all $\rho$ with p-values above $0.32$,
but may contain states with p-values as low as $0.23$, because the
true value of $D(\rho|\{x_i\})$ can be as large as
$2[L(\rho_k|\{x_i\})+r_k-L(\rho|\{x_i\})] = 105.04+(2\times2)$ for
those states.  Note that the p-values included in the region are not
data dependent provided we choose the stopping rule beforehand.  If we
stop at $r_k\precsim 1.5$ and set the significance level at $0.05$,
corresponding to a threshold $t=123.22$, the confidence region may
contain states with p-values as low as $0.03$.

\section{Expectation value confidence interval stopping rule}

Another way to utilize tomographic data is to estimate expectation
values, such as $\Tr(\rho_{\mathrm{true}} A)$ and give a confidence
interval for the estimate at a given level of significance $s$.  Let
$F(f)=\{\rho:\Tr(\rho A)=f\}$ be a level set for
$\Tr(\rho_{\mathrm{true}} A)$.  The dimensionality of this level set
is $d^2-2$. Let $\phi_{\mathrm{ML},f}$ be the state in $F(f)$
maximizing the likelihood.  To establish a confidence region for $f$
via a likelihood-ratio test, we use the statistic
$D(\phi_{\mathrm{ML},f}|X)$.  For each $f$ there is an associated
p-value (the p-value of the log-likelihood difference between $\rhoml$
and $\phi_{\mathrm{ML},f}$), and all $f$'s with p-value at least $s$
are in the confidence region.  By Wilks's Theorem, the statistic
$D(\phi_{\mathrm{ML},f}|X)$ has distribution $\chi^2(1)$.  Let $t$ be
the maximum value of $D(\phi_{\mathrm{ML},f}|X)$ for which $f$ is a
member of the confidence interval at significance level $s$.
Following the discussion above, $t$ is related to $s$ through the
integral of the $\chi^2(1)$ distribution.  The confidence region for
$f$ is $C = \{f : D(\phi_{\mathrm{ML}}|\setx) \leq t\}$.  It is
necessary to adapt the stopping rule to the maximum-likelihood problem
constrained to $F(f)$.  It is not practical to compute
$L(\phi_{\mathrm{ML},f})$ for all $f$, neither is it necessary to do
so. We may use the Lagrange multiplier technique to compute
$L(\phi_{\mathrm{ML},f})$.  With $\lambda$ as the Lagrange multiplier,
we maximize $K(\rho,\lambda)=L(\rho)+\lambda \Tr(\rho A)$.  This
function is still concave over the full space of density matrices and
can be maximized by the same methods as $L(\rho)$ after replacing
$R(\rho)$ with $R(\rho)+\lambda A$.  In the standard Lagrange
multiplier technique, one usually solves an equation for the value of
$\lambda$ that corresponds to the desired constraint $f$.  Solving
such an equation in this case would be difficult, and we do not know
the desired $f$ in advance.  We need to approximate the values of $f$
that are the limits of the confidence interval.  To accomplish this,
we choose a value for $\lambda$ and maximize $K(\rho,\lambda)$ to find
a state $\phi_{\mathrm{ML},\lambda}$ that is the maximum-likelihood
state obeying the constraint
$\Tr(\phi_{\mathrm{ML},\lambda}A)=f_\lambda$, where $f_\lambda$
depends on the choice for $\lambda$.  If $\phi_{\mathrm{ML},\lambda}$
has the desired log-likelihood difference $t$, $f_\lambda$ marks one
boundary of the confidence interval.  If not, we search for the
desired $\lambda$ by re-maximizing $K(\rho,\lambda)$ with different
choices of $\lambda$. This search is simplified by the observation
that $\lambda$ is monotonically related to the log-likelihood of
$\phi_{\mathrm{ML},\lambda}$.  This follows from concavity of the
log-likelihood: The maximum log-likelihood $L(f)$ on level set $F(f)$
is a concave function of $f$ and $-\lambda$ is the slope of $L(f)$ at
$f=f_\lambda$.

Given an iterative method for maximizing $K(\rho,\lambda)$, the upper
bound on log-likelihood derived from $r_k$ generalizes, yielding a
bound $r(\phi_j)$ on the maximum possible increase in
$K(\rho,\lambda)$ at the $j$'th iterate $\phi_j$. Let $f_j=\Tr(\phi_j
A)$.  Since $K(\rho,\lambda)$ is constant on level sets $F(f)$,
$r(\phi_j)$ is a bound on $L(\phi_{\mathrm{ML},f_j})-L(\phi_j)$.
Given the iterate found after stopping, we can bound the true value of
the desired log-likelihood difference by
\begin{eqnarray*} D(\phi_{\mathrm{ML},f_j}|\{x_i\}) \geq
D_{\mathrm{lb}} =
2\left[L(\rho_k|\{x_i\})-L(\phi_j|\{x_i\})-r(\phi_j)\right], \\
D(\phi_{\mathrm{ML},f_j}|\{x_i\}) \leq D_{\mathrm{ub}} =
2\left[L(\rho_k|\{x_i\})+r(\rho_k)-L(\phi_j|\{x_i\})\right].
\end{eqnarray*} 
For a conservative approximation of the desired confidence interval,
we run the iterative method with a stopping rule, seeking lower and
upper bounds $f_j$ for which $D_{\mathrm{lb}}$ is close to $t$. To
ensure a conservative estimate, it should be at least $t$.  To avoid
an unnecessarily large confidence interval, we should ensure that
$r(\rho_k)$ and $r(\phi_j)$ are sufficiently small fractions of
$\sqrt{2}$, the standard deviation of $\chi^2(1)$.  For example,
suppose that we wish to approximate a confidence interval at
significance level $0.32$. The threshold for
$D(\phi_{\mathrm{ML},f_j})$ is $t=0.99$. Suppose that we stop
iterations when $r(\rho_k),r(\phi_j)\precsim 0.3$ and set the
confidence interval according to $\{f:D_{\mathrm{lb}}\leq t\}$.  Then
the confidence interval includes all $f$'s with p-values larger than
$0.32$ and may contain $f$'s with p-values as low as $0.21$.  If we
set the threshold at $t=3.84$ according to a significance level of
$0.05$ and stop at $r_k,r(\phi_j)\precsim 0.2$, the confidence
interval may contain $f$'s with p-values as low as $0.04$.

\section{Numerical simulation}

\begin{figure}
\includegraphics[width=\textwidth]{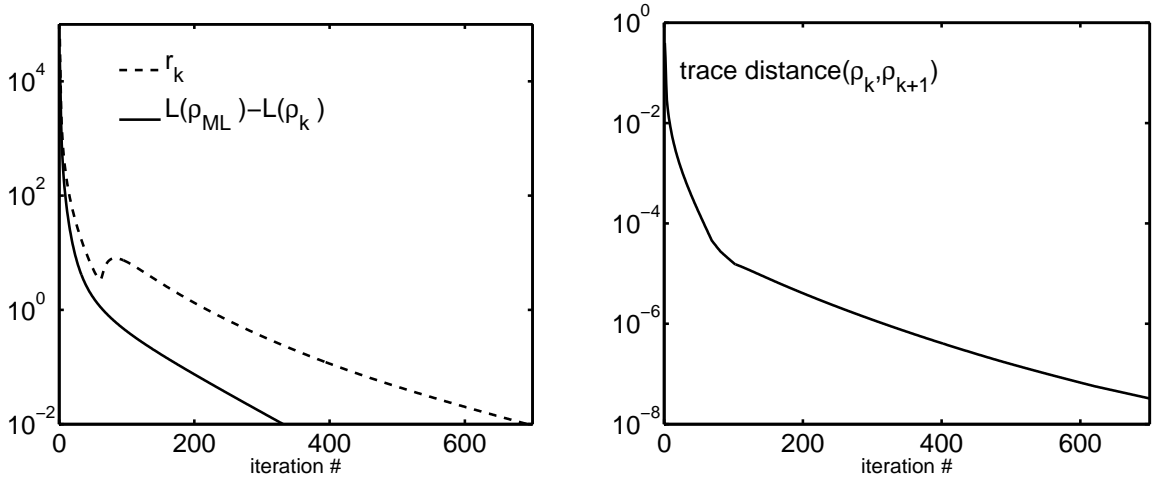}
\caption{The left graph shows $L(\rhoml)-L(\rho_k)$ and $r_k$ as a
  function of iteration number $k$ for optical homodyne tomography of
  a $d=11$ (10 photon) dimensional quantum state. The right graph
  shows the trace distance between state $\rho_k$ and
  $\rho_{k+1}$. Trace distance was calculated as
  $\Tr[|\rho_k-\rho_{k+1}|]/2$.
\label{r_vs_iteration}}
\end{figure}

To illustrate the behaviour of the bound used for the stopping rules,
we simulated homodyne measurements \cite{lvovsky2005} of a state
created by sending a superposition of optical coherent states
($|\alpha\rangle+|-\alpha\rangle$, unnormalized, with $\alpha=1$)
through a medium whose transmissivity is 80 \%.  The homodyne
measurements are 90 \% efficient.  The Hilbert space was truncated at
10 photons.  Results are shown in Fig.~\ref{r_vs_iteration}, where we
have used the $R\rho R$ algorithm to maximize the likelihood.  To make
this figure, we computed 1122 iterations and assigned $\rhoml =
\rho_{1122}$.  Further iterations suffered from numerical errors.
After an initial phase of very fast likelihood increase, the
convergence rate significantly drops.  As expected,
$L(\rhoml)-L(\rho_k)$ decreases with each iteration, but $r_k$
sometimes increases.  There is a significant gap between $r_k$ and
$L(\rhoml)-L(\rho_k)$, so it would be helpful to find tighter bounds
to prevent unnecessary iterations.  Perhaps the
bound could be made more tight using a higher order expansion of the
log-likelihood function in $\epsilon$.  Without a reliable stopping rule
such as one based on $r_k$, a simple strategy is to stop iterations
when the difference between successive density matrices obtained is
very small.  For comparison, the figure includes a plot of the trace
distance between $\rho_k$ and $\rho_{k+1}$.  According to the rough
guidelines given above, if we want to use the result of this
computation to obtain a confidence interval for an expectation value
of the true state, we might halt iterations when $r_k \leq 0.1$, at
which point the trace distance between $\rho_k$ and $\rho_{k+1}$ is
$3.6\times 10^{-7}$.  In general, the relationship between $r_k$ and
trace distance is dependent on the situation.

\section{Conclusion}

Our bounds on the likelihood ratios hold regardless of Wilks's
Theorem, but we have used Wilks's Theorem to construct the confidence
regions described above.  Wilks's Theorem must be applied carefully
(if at all) when performing quantum state tomography.  In particular
the techniques discussed above cannot be used if the true state has
eigenvalues that are statistically close to $0$ or if there is
insufficient data for the limiting distributions to be good
approximations.  Both of these situations are common in applications
of tomography and can result in bad confidence regions and excessive
biases.  For example, we encountered such difficulties analyzing the
data reported in Ref.  \cite{mallet2011}; see the discussion in this
reference's supplementary materials.  When Wilks's Theorem cannot be
applied, one must resort to other techniques such as the robust bounds
on log-likelihood differences described in \cite{christandl2011,
  blume-kohout2012} or parametric or non-parametric
bootstrap~\cite{efron1993} for estimating statistical errors and
confidence regions.  The bootstrap methods require running
maximum-likelihood algorithms on many simulated or resampled data
sets.  Judicious use of one of the stopping rules given above can
significantly reduce the number of iterations required when optimizing
the likelihood, thereby making it possible to implement bootstrap with
more resampled data sets to obtain better estimates. However, if bias
in the maximum-likelihood estimate is large, confidence regions
constructed by bootstrap may also be unreliable \cite{efron1993}.

We have presented an upper bound $r_k$ on the log-likelihood
difference of the maximum-likelihood state and the currently found
state in iterative algorithms for maximum-likelihood tomography.  The
bound is easily computed from the gradient of the log-likelihood
function and can be used in stopping rules for confidence regions or
decisions that use the likelihood ratio as a test statistic.

\ack

We thank Robin Blume-Kohout, Kevin Coakley, Mike Mullan, and Yanbao
Zhang for helpful discussion.  We are grateful to NSF and NIST for the
award of a SURF fellowship to M. Girard.  Portions of this article are
official contributions of the National Institute of Standards and
Technology and are not subject to copyright in the United States.

\section*{References}
\bibliography{journal_titles_abbreviated,bib}

\end{document}